\documentclass[structabstract]{aa} 

\usepackage{epsfig}
\usepackage{graphicx}
\usepackage{txfonts}
\usepackage{amssymb}
\usepackage{natbib}                               
\bibpunct{(}{)}{;}{a}{}{,}

\def\compps{{\sc compps}} 
\def\taut{\tau_{\rm T}}  
\def\bb{{\sc bb}} 
\def\nh{{$N_{\rm H}$}} 
\def\Integ{{\em INTEGRAL}} 
\def\rxte{{\em RXTE}} 
\def\swift{{\em Swift}}

\def\igr{IGR~J17498--2921}

\def\be{\begin{equation}} 
\def\ee{\end{equation}} 
 
\begin{document} 
\title{Spectral and timing properties of  the accreting  X-ray millisecond pulsar IGR~J17498--2921} 
 
\author{M. Falanga\inst{1}
\and L. Kuiper\inst{2} 
\and  J. Poutanen\inst{3} 
\and D. K. Galloway\inst{4}
\and E. Bozzo\inst{5}
\and A. Goldwurm\inst{6,7}  
\and W. Hermsen\inst{2,8} 
\and L. Stella\inst{9}           
} 
 
 
\titlerunning{Accreting  X-ray millisecond pulsar IGR~J17498--2921}  
\authorrunning{Falanga et al.}  
  
\institute{International Space Science Institute (ISSI), Hallerstrasse 6, CH-3012 Bern, Switzerland 
 \email{mfalanga@issibern.ch}
\and SRON--Netherlands Institute for Space Research, Sorbonnelaan 2, 3584 CA, Utrecht, The Netherlands 
\and Astronomy Division,  Department of Physics, P.O. Box 3000, FI-90014 University of Oulu, Finland  
\and Monash Center for Astrophysics, School of Physics, and School of Mathematical Sciences, Monash University, VIC 3800, Australia
\and ISDC, Data centre for astrophysics, University of Geneva, Chemin d'\'Ecogia 16, 1290 Versoix, Switzerland
\and Service dÕAstrophysique (SAp), IRFU/DSM/CEA-Saclay, 91191 Gif-sur-Yvette Cedex, France 
\and Unit\'e mixte de recherche Astroparticule et Cosmologie, 10 rue Alice Domon et Leonie Duquet, F-75205 Paris, France  
\and Astronomical Institute ``Anton Pannekoek'', University of Amsterdam, Science Park 904, 1098 XH, Amsterdam, The Netherlands 
\and INAF--Osservatorio Astronomico di Roma, via Frascati 33, 00040 Monteporzio Catone (Roma), Italy  
             } 
 
   \date{ } 
 
  
  \abstract 
{\igr\ is the third  X-ray transient accreting millisecond pulsar discovered by \Integ. It was in outburst for about 40 days beginning on August 08, 2011.} 
{We analyze the spectral and timing properties of the object and the characteristics of 
X-ray bursts to constrain the physical processes responsible for the X-ray production in 
this class of sources. 
}  
  {We studied the broad-band spectrum of the persistent emission in the 0.6--300 keV energy band
  using simultaneous \Integ, \rxte, and \swift\ data obtained in August-September 2011.
    We also describe  the timing properties in the 2--100 keV energy range  
    such as the outburst lightcurve, pulse profile, pulsed fraction, pulsed emission, time lags, 
    and study the properties of X-ray bursts discovered by \rxte, \swift, and \Integ\ and the recurrence time.}
  {The  broad-band average spectrum is well-described by thermal Comptonization with an electron temperature of
  $kT_{\rm e}\sim 50$ keV, soft seed photons of  $kT_{\rm bb}\sim 1$ keV, and Thomson optical 
  depth $\taut\sim1$ in a slab geometry.  The slab area corresponds to a black body radius of $R_{\rm bb}\sim9$ km. During the outburst, the spectrum stays remarkably stable with plasma and soft seed photon
temperatures and scattering optical depth that are constant within the errors.  This behavior has been interpreted as indicating that the X-ray emission originates above the neutron star (NS) surface in a hot slab (either the heated NS surface or the accretion shock). 
The \Integ,  \rxte, and \swift\ data reveal the X-ray  pulsation at a period of 2.5 milliseconds up to $\sim$65 keV. The pulsed fraction is consistent with being constant, i.e. energy independent and has a typical value of 6--7\%. The nearly sinusoidal pulses show soft lags that seem to saturate near 10 keV at a rather small value of $\sim -60\mu$s  with those observed in other accreting pulsars. 
The short burst profiles indicate that there is a hydrogen-poor material at ignition, which suggests either that the accreted material is hydrogen-deficient, or that the CNO metallicity is up to a factor of about two times solar. However, the variation in the burst recurrence time as a function of $\dot{m}$ (inferred from the X-ray flux) is much smaller than predicted by helium-ignition models.} 
 {}
 
\maketitle

\section{Introduction} 
\label{sec:intro}

\igr\ was discovered with \Integ\ on August 11, 2011 \citep{Gibaud11}. 
Pulsations  at a frequency  of $\sim$401 Hz \citep{Papitto11ATel3556} observed from the source, 
make it  the fourteenth known accreting  millisecond X-ray pulsar (AMXP). 
The orbital period is about 3.8 hours \citep{Markwardt11ATel3561}. 
\igr\ also shows X-ray bursts \citep{Ferrigno11ATel3560}.  
The source is located at  
$\alpha_{\rm J2000} = 17^{\rm h}49^{\rm m}55\fs35$ and $\delta_{\rm J2000} = -29{\degr}19\arcmin19\farcs6$, with an associated uncertainty of $0\farcs6$ at the 90\% c.l. \citep{Bozzo11ATel3558,Chakrabarty11ATel3606}, at a most likely distance of $\sim 7.6$ kpc \citep{Linares11ATel3568}.
 
Using {\rm Chandra} archival data, \igr\ has also been observed in quiescence at a luminosity level of $\sim 2\times10^{32}$ erg cm$^{-2}$ s$^{-1}$ 
\citep[for the source distance of 8 kpc, see][]{Jonker11ATel3559}. The near IR and optical counterpart has also been found to be very faint 
\citep{Greiss11ATel3562,Russell11ATel3622,Torres11ATel3638,vandenBerg11ATel3634}. More details about this source can be found in \citet{Papitto11}. 
  
The AMXPs are rapidly spinning, old, recycled neutron stars (NSs) hosted in low-mass X-ray binaries. 
For reviews of the properties of these objects, we refer the reader to the papers by \citet{Wijnands2010} and \citet{Poutanen06}.   
 
In this paper, we report on \Integ, \swift, and Rossi X-ray Timing Explorer (\rxte) observations of \igr\ during its August--September 2011 outburst, as well as simultaneous observations by the same satellites during the period August 17--20, 2011 (modified Julian date MJD 55790--55793). 
We study the lightcurves, broad-band spectra, outburst spectral evolution, and timing properties of the source. 
The properties of the largest set of X-ray bursts from this source are investigated as well.

\section{Observations and data} 
 
\subsection{INTEGRAL} 
\label{sec:integral}  
 
We analyzed all the \Integ\ \citep{Winkler03} pointings available in the direction of the source. The observations during  
satellite revolutions 1078--1083 started on August 11 (MJD 55784.948) and ended on August 27 (MJD 55800.051). 
This also includes a target of opportunity observation during satellite revolution 1080, beginning on August 17 (MJD 55790.9242) and ending on August 20, 2011 (MJD 55793.5631), with a total net exposure time of 214 ks.  
The data from the IBIS/ISGRI coded mask telescope \citep{Ubertini2003,Lebrun2003} consist of 168 stable pointings with a source position offset $\lesssim 12\fdg0$ from the center of the field of view with a total net exposure time of $\sim$600 ks.
The data from the JEM-X monitor modules 1 and 2 \citep{Lund03} consist of 92 stable pointings for a total net exposure time of $\sim$323 ks, with a source position offset of $<3\fdg5$ from the center of the field of view. 
The data reduction was performed using the standard Offline Science Analysis (OSA)\footnote{http://www.isdc.unige.ch/integral/analysis} software version 9.0. The algorithms used for the spatial and spectral analysis are described in \citet{Goldwurm03}.

\begin{figure}
\centering 
\includegraphics[width= 0.90\columnwidth]{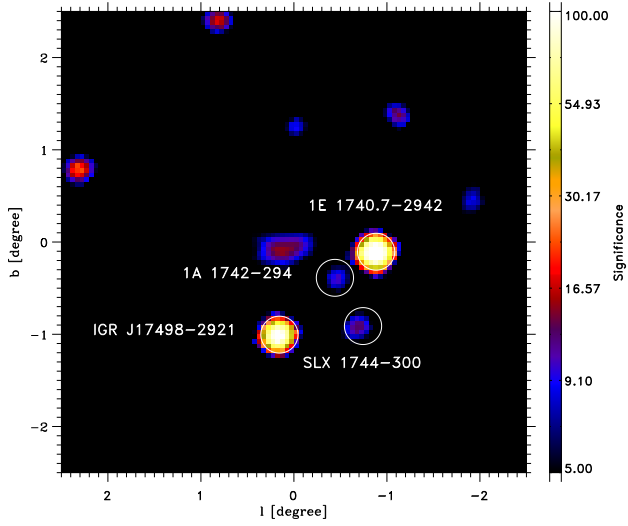}
{\caption[]{{\it INTEGRAL}/ISGRI sky image of the field of view around \igr\
 in the 20--100 keV band for an effective exposure of 210 ks. The size of each pixel in the image corresponds to 3$\arcmin$. 
As the $\sim1\degr$-radius {\it RXTE}/PCA field of view also includes 1A 1742-294 ($0\fdg86$ from IGR~J17498$-$2921) and SLX~1744$-$300 ($0\fdg9$), the data from that instrument also has contributions from those sources, as well as the diffuse emission from the Galactic center, in particular the non-imaging PCA instrument was unable to associate the detected bursts with the three burster sources.
\label{fig:fig1}}}
\end{figure} 

In Fig.~\ref{fig:fig1}, we show part of the ISGRI field of view (significance map) around the position of \igr\ (in the 20--100 keV energy range).  
The source is detected at a significance level of $\sim113\sigma$. 
The determined position is $\alpha_{\rm J2000} = 17^{\rm h}49^{\rm
m}56\fs02$ and $\delta_{\rm J2000} = -29{\degr}19\arcmin20\farcs7$, with an associated uncertainty
of $0\farcm4$ at the 90\% confidence level. 
The offset with respect to the {\em Chandra} position \citep{Chakrabarty11ATel3606} is  $0\farcm15$.

In the same figure, we also indicate with white circles the position of two burster sources SLX~1744--300/299 and 1A~1742--294, as well as the  black hole candidate 1E~1740.7--2942. These sources are detected in the total \Integ/ISGRI mosaic at $\sim$13$\sigma$, $\sim$11$\sigma$, and $\sim$151$\sigma$ and located at $0\fdg9$, $0\fdg86$, and $1\fdg37$, respectively, from the \igr\ position (see Fig.~\ref{fig:fig1}). 
Using a simple absorbed power-law model, we determined of the broad-band JEM-X/ISGRI spectra  the unabsorbed flux to be $\sim$1.6$\times$10$^{-10}$, $\sim4.4\times10^{-10}$, and $\sim6.3\times10^{-10}$ erg cm$^{-2}$ s$^{-1}$ in the 3--20 keV band, of SLX~1744--300/299, 1A~1742--294, and 1E~1740.7--2942, respectively. 
We note that in the JEM-X data we clearly distinguish the type-I X-ray bursts emitted by \igr\ from the bursts 
emitted by SLX~1744--300/299 and 1A~1742--294 (see Sect. \ref{sec:rxte}).

\subsection{RXTE and Swift} 
\label{sec:rxte}  
  
We used all the publicly available data from the Proportional Counter Array
(PCA, 2--60 keV; \citealp{Jahoda96}) on-board the \rxte\ satellite and the  \swift\ X-Ray Telescope (XRT) \citep{Burrows05} data (0.6--10 keV) in window-timing mode. \igr\ was observed with \swift\ and \rxte\  between August 12 and September 22, 2011 (MJD
55786.1--55826.4) for total net exposure times of $\sim48.3$ ks and $\sim 94.2$ ks, respectively. 
The same dataset was first published by  \citet{Papitto11}, hence we refer the reader to this paper for the \swift/XRT and \rxte/PCA data reduction procedure. However, Fig.~\ref{fig:fig1} shows that the angular separation between different sources, e.g., SLX~1744--300/299 and 1A~1742--294, is smaller than the field of view of the non-imaging instruments on-board \rxte\
($\sim1\degr$), thus we were unable to separate the contributions of these objects to the total X-ray flux.  
In Sect. \ref{sec:integral}, we report  from \Integ\ observations the fluxes of the sources, SLX~1744--300/299 and 1A~1742--294 (and 1E~1740.7--2942), 
in the same energy band used for \rxte/PCA data. 

\begin{figure} 
\centering 
\centerline{\epsfig{file=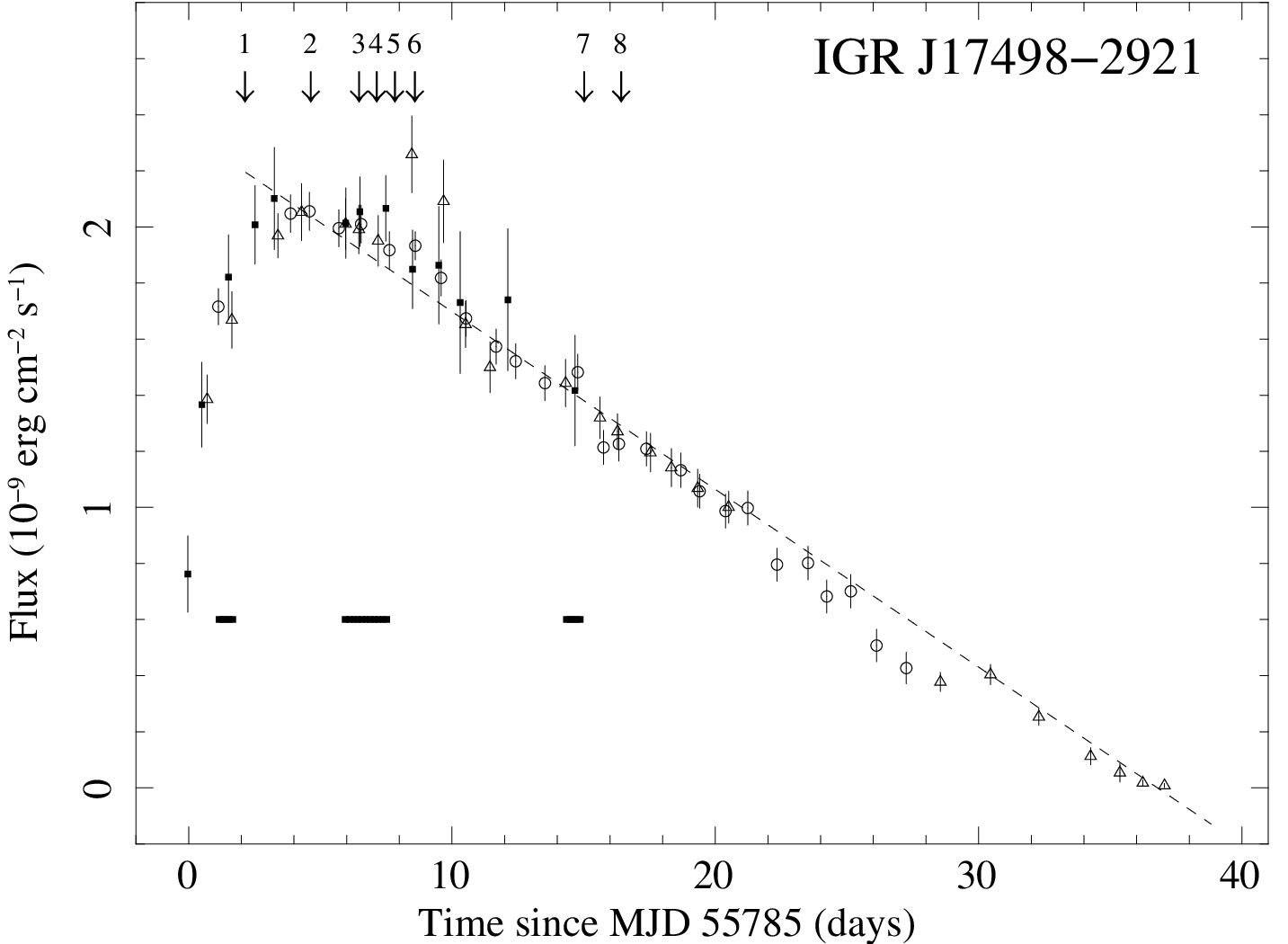,width=6.5cm,angle=-90} } 
\caption{\rxte/PCA (open circles), \swift/XRT (open triangles), and \Integ/ISGRI (filled squares) outburst lightcurve of \igr. 
For plotting purpose, we chose a time bin  of one day. The count rate was converted into flux (0.1--300 keV) using the
spectral results reported in Sect. \ref{sec:spec}. The arrows indicate the times of the detected
X-ray bursts (see Table~\ref{tab:burst}). The dashed lined correspond to the best-fit linear decay 
$F_{-9}=  -0.64\, t_{\rm day}\pm0.09$. 
The three horizontal bars correspond to the time intervals of our spectral results reported in 
Table~\ref{tab:table1}.
}
\label{fig:fig2} 
\end{figure}

\subsection{Outburst lightcurve}
\label{sec:lc} 

The outburst rise of \igr\ was observed from an earlier time, and with a better coverage, than for any previous  AMXP.
The rise time was shorter than two days. The flux stayed almost constant for five days and  decayed 
afterwards linearly to the quiescence level. The outburst profile shape was similar to those of other AMXPs, e.g., SAX J1808.4--3658 \citep{Gilfanov1998}, 
XTE J1751--305 \citep{Gierlinski05}, or IGR J17511--3057 \citep{Falanga11},  and the outburst lasted in total about six weeks. 
However, since the \rxte/PCA data were contaminated by other sources in the field of view, 
it is impossible to accurately follow the lightcurve decay, i.e. it is unclear whether the lightcurve decayed exponentially
until it reached a break and that the flux then dropped linearly to the
quiescence level as observed in other sources \citep[see e.g.,][and references therein]{Falanga11}. 
The break may have occurred in the lightcurve around day 30 (see Fig.~\ref{fig:fig2}). 
However, a statistically acceptable fit was a linear relation over the whole outburst as shown in Fig.~\ref{fig:fig2}. 
The lightcurves of AMXPs during their outburst  have been modeled in the past using various models \citep[see e.g.,][]{Powell07,Hartman11}.

The times of the detected type-I X-ray bursts are indicated with arrows in Fig.~\ref{fig:fig2}. 
\rxte\ has detected two, \Integ/JEM-X five, and \swift\ two type-I X-ray bursts. 
The burst number 3 was observed simultaneously by \Integ\ and \swift. 
The burst properties are discussed in detail in Sect. \ref{sec:burst}.

\section{Spectral analysis of persistent emission} 
\label{sec:spectra}
  
The analysis of the persistent emission spectra was carried out using {\sc xspec} version 12.7.0. 
For the contemporaneous data, we combined the low-energy range 0.6--10 keV \swift\ data, the 3--22 keV \rxte/PCA data, and the 5--300 keV \Integ/JEM-X/ISGRI data. For each instrument, a multiplication factor was included in the fit to take into account the
 uncertainty in the cross-calibration of the instruments. 
This factor was fixed at one for the ISGRI data.  
To follow the outburst spectra outside the \Integ\ observation interval, we used the \rxte/PCA and \swift/XRT data.  To take into account the contamination by the Galactic ridge emission and emission from the different sources in the \rxte/PCA field of view, we used as background the {\em RXTE}/PCA data collected when the pulsar, \igr, was no longer detected \citep[see also][]{Papitto11}. All uncertainties in the single spectral parameters are given at a 90\% confidence level.
   
\begin{figure} 
\centering 
\centerline{\epsfig{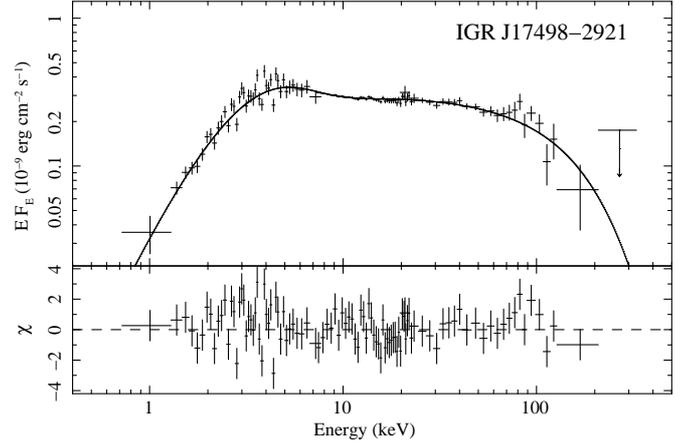} } 
\caption{Unfolded absorbed broad-band spectrum of \igr\ 
collected during the peak of the outburst, interval MJD 55790.948--55792.523, and the best-fit thermal Comptonization model \compps. 
The data points are from the \swift/XRT (0.6--8 keV), \rxte/PCA (3--22 keV), and \Integ/ISGRI (20--300 keV) instruments,
respectively. The total model spectrum is shown by a solid line. The lower
panel shows the residuals between the data and the model.
}
\label{fig:fig3} 
\end{figure} 
 
\subsection{Broad-band spectrum} 
\label{sec:spec} 
 
 We studied in detail the broad-band X-ray spectrum of \igr\ in the energy range 0.6--300 keV 
 during the rise, peak, and decay of the outburst using the joint \Integ/ISGRI, \rxte/PCA, and \swift/XRT data.
 For all the data, we removed the time intervals corresponding to
 the bursts.  We fitted each joint XRT/PCA/ISGRI  
 spectrum with a thermal Comptonization model, \compps, in the slab geometry \citep{Poutanen96}. 
The same model was used to fit the broad-band  spectra of previously observed
AMSPs \citep[e.g.,][]{Gierlinski05,Falanga05a,Falanga05b,Falanga07,Ibragimov09,Falanga11}. 
The main model parameters are the Thomson optical depth 
$\tau_{\rm T}$ across the slab, the electron temperature $T_{\rm e}$,
the temperature $T_{\rm bb}$ of the soft-seed blackbody photons assumed to be
injected at the bottom of the slab, the emission area $A_{\rm bb}$ corresponding to this blackbody, 
and the inclination angle $\theta$ between the slab normal and the line of sight. The best-fit
spectral parameters during the different outburst epochs are reported in Table
\ref{tab:table1}. In Fig.~\ref{fig:fig3}, we show the unfolded and absorbed spectra collected 
during the peak of the outburst over the interval MJD 55790.948--55792.523 and the best-fit  {\sc compps}  model. 
In these fits, the normalizations of the XRT and PCA data relative to ISGRI  were $\sim$0.7 and 1.1, respectively. 
    
\begin{table}
\caption{Best-fit spectral parameters 
of the \compps\ model to the XRT/PCA/ISGRI data}
\centering
\begin{tabular}{llll} 
\hline 
Parameters & Rise & Peak & Decay\\
\tiny{Time interval ($+$55785 MJD)} & \tiny{1.157--1.682}&\tiny{5.948--7.523} &\tiny{14.348--14.873}\\
\hline
\noalign{\smallskip}  
$N_{\rm H} (10^{22} {\rm cm}^{-2})$ & $1.5_{-0.2}^{+0.2}$& $1.2_{-0.16}^{+0.14}$&$1.2_{-0.20}^{+0.23}$\\ 
$kT_{\rm e}$ (keV)& $59.8_{-6.0}^{+6.5}$ & $52.8^{+4.4}_{-4.2}$&$33.2^{+6.5}_{-6.3}$\\ 
$kT_{\rm bb}$ (keV)& $1.03_{-0.07}^{+0.07}$&$0.98^{+0.05}_{-0.05}$&$0.96^{+0.09}_{-0.09}$\\ 
$\tau_{\rm T}$ &  $0.8_{-0.07}^{+0.08}$&$0.9^{+0.08}_{-0.07}$&$1.2^{+0.17}_{-0.24}$\\ 
$R_{\rm bb}$\tablefootmark{a} (km)& $8.2^{+1.4}_{-1.2}$& $9.9^{+1.2}_{-1.0}$&$8.2^{+2.6}_{-2}$ \\ 
$\cos \theta $ & $0.48^{+0.02}_{-0.01}$& $0.51^{+0.02}_{-0.01}$&$0.52^{+0.03}_{-0.03}$\\
$\chi^{2}/{\rm dof}$ & 168/126 & 288/263 &178/199\\
$F_{\rm bol}$\tablefootmark{b} ($10^{-9}$ erg cm$^{-2}$ s$^{-1}$) &1.7$\pm$0.2& 2.0$\pm$0.2& 1.4$\pm$0.2\\
\noalign{\smallskip}  
\hline  
\noalign{\smallskip}  
\end{tabular}  
\tablefoot{ \tablefoottext{a}{Assuming a source distance of 8 kpc.}
\tablefoottext{b}{Unabsorbed flux in the 0.1--300 keV energy range.}
}
\label{tab:table1} 
\end{table} 

We also studied the spectral evolution (rebinned over one day) during the whole outburst using contemporaneous 
 \rxte/PCA, \swift/XRT, and  \Integ/ISGRI (0.8--300 keV) spectra, and the combined or single \rxte/PCA (3--22 keV) and 
 \swift/XRT (0.6--10 keV) observations. 

 We found that the spectral evolution of the outburst is marked by a nearly constant plasma temperature $kT_{\rm e}\sim55$ keV, soft-seed photons emission $kT_{\rm bb}\sim1$ keV, and the optical depth $\tau_{\rm T}\sim1$. 
No statistically significant variations were measured. 
Similar stabilities in the spectral shape at different fluxes has been observed for a number of 
other AMSPs \citep{Gilfanov1998,Gierlinski05,Falanga05b,Poutanen06,Ibragimov09,Ibragimov11,Falanga11}.
This behavior has been interpreted as evidence that the X-ray emission originates above the NS surface 
in a hot slab (either the heated NS surface or the accretion shock). 
The energy dissipation takes place there and the electron temperature is determined 
by the flux of soft seed photons produced by the reprocessing of the hard X-ray radiation 
at the NS surface \citep[see e.g.][]{HM93,SPS95,Poutanen96,MBP01}. 

The broad-band spectrum has a clear signature of thermal emission below $\sim$7 keV (see  Fig.~\ref{fig:fig3}). 
Thermal emission can in principle originate from the cool, truncated accretion disk, as observed e.g. in SAX J1808.4--3658 \citep{Patruno09,Kajava11} and IGR~J17511--3057 \citep{Papitto10,Ibragimov11}. 
However, it normally contributes only below 3 keV.
In addition, the rather constant pulsation amplitude as a function of energy  (see Sect. \ref{sec:pulsefrac}) does not 
support this interpretation. 
Thus, this thermal emission is associated with the heated NS surface beneath and around the accretion hotspot.

\section{Timing characteristics}
\label{sec:tmchar}

\subsection{Data}
\label{sec:time_data}

The emission of \igr\ was studied in detail using data from \rxte/PCA (3--36 keV), HEXTE (15--250 keV) and \Integ/ISGRI (20--300 keV). Because the flux decays over the course of the outburst, we selected only data from these instruments where the observations had overlapped in time in order 
to minimize systematic (pulsed) flux offsets between the measurements. This selection yielded the following interval for the overlapping data period from MJD 55786.114 (first RXTE observation; 96435-01-01-00) to 55800.061 (last \Integ\ observation; Rev. 1083 Scw. 100010).

We adopted standard selection criteria for the PCA screening process, ignoring time intervals in which bursts and detector breakdowns occur, and obtained the exposure times for this data period for the PCU-detectors of 0--4,11.98 ks, 15.74 ks, 73.70 ks, 10.24 ks, and 23.75 ks, respectively. The data were collected in {\tt GoodXenon} mode allowing high-time resolution (0.9$\mu$s) analyses in 256 spectral bins. In the timing analysis, data from all PCU detection layers were used to ensure sensitivity to photons with energies in excess of $\sim 10$ keV.

For HEXTE, we used the cluster-0 data, because only this cluster had operated in "on-source" staring mode, since July 13, 2006, and we obtained a dead-time corrected exposure of 50.29 ks.

The ISGRI data, collected during \Integ\ revolutions 1078--1083 with a start science window 310010 of \Integ\ revolution 1078, were screened for the effects caused by solar flare activity and Earth radiation belt passages, but none of the pointings showed evidence of any effects. Time periods in which burst events occurred from any source in the ISGRI field of view were excluded from further analysis. 
This screening process yielded a ``good'' exposure time of 392.4 ks.
Furthermore, in the timing analysis we selected only time stamps of events with rise times between channels 7 and 90 \citep{Lebrun2003} that had been detected in non-noisy detector pixels with an illumination factor of more than 25\%.

The selected time stamps of all instruments used in the timing study were converted to arrival times at the solar system barycenter by taking into account the orbital motion of the spacecraft and correcting for acceleration effects along the binary orbit. In this process we used the position of the X-ray counterpart to \igr\ reported by \citet{Chakrabarty11ATel3606}.

\begin{figure}
\centering 
\includegraphics[width= 0.85\columnwidth]{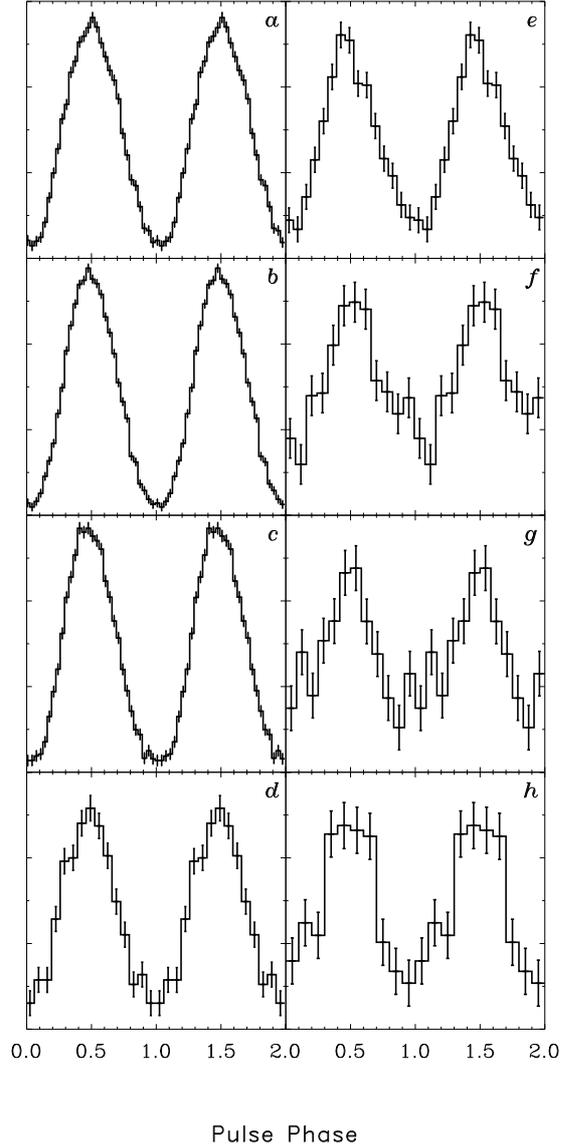}
              {\caption{Pulse-profile collage of \igr\ using data from \rxte/PCA, HEXTE, and \Integ/ISGRI. Panels a--d (PCA) correspond to the energy intervals 1.6--4.1, 4.1--8.7, 8.7--18.4, and 18.4--37.0 keV, respectively. Panels e--f show the HEXTE pulse profiles in the bands 15.6--31.0 keV and 31.0--65.1 and
              in panels g--h the ISGRI profiles for the bands 20--30 keV and 30--65 keV.  The error bars represent $1\sigma$ statistical errors. All profiles reach their maximum near phase $\sim 0.5$. 
              The y-axis is given in units of counts per bin.
              \label{fig:profcol}}}
\end{figure}

\subsection{Pulse profiles and time lags}

We folded the barycentered arrival times using the ephemeris given in \citet{Markwardt11ATel3561}. However, we used a different time for the epoch of the ascending node of MJD 55786.18099710 (TDB), because the one reported in \citet{Markwardt11ATel3561} actually refers to $T_{\pi/2}$.

Panels a--d of Fig.~\ref{fig:profcol} show the PCA pulse profiles for the energy intervals,
1.6--4.1, 4.1--8.7, 8.7--18.4, and 18.4--37.0 keV, respectively. In panels e--f, the HEXTE profiles are shown for the bands 15.6--31.0 keV and 31.0--65.1 keV, and panel g--h show the ISGRI profiles for the energy intervals 20--30 keV and 30--65 keV. Above $\sim 65$ keV, no significant pulsed emission can be detected in either HEXTE or ISGRI data.
The mutual alignment of the \rxte/PCA, HEXTE, and \Integ/ISGRI profiles within the equivalent energy bands is better than $50\mu$s.

\begin{figure} 
\centering 
\includegraphics[width= 0.85\columnwidth]{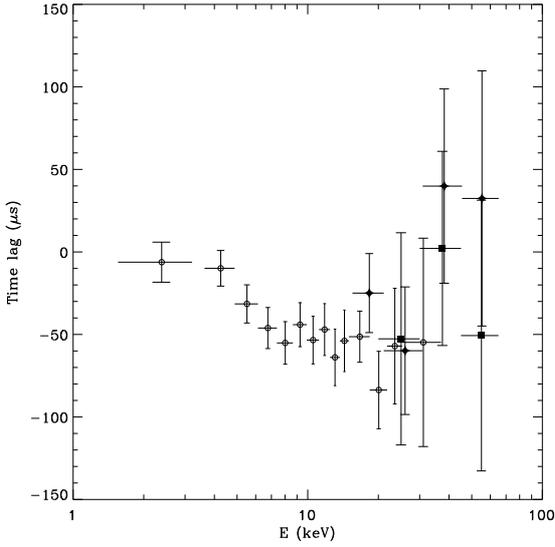}
\caption{Time lag as a function of energy in the $\sim$2--100 keV
energy range combining \rxte/PCA (1.6--37.0 keV; open circles),
\rxte/HEXTE (15--65 keV; filled diamonds), and \Integ/ISGRI
(20--65 keV; filled squares) measurements.}
\label{fig:timelag} 
\end{figure}

The global arrival times of the pulses as a function of energy were studied using the PCA profile for PHA channel-9 
($\sim$3.2--3.6) keV as a reference template. Cross-correlation of the pulse profiles obtained for other energy bands with this template yielded the time lags shown in Fig.~\ref{fig:timelag}. Beyond $\sim$4 keV, a declining trend sets in, i.e.  
the hard X-ray photons arrive earlier than the soft ones.  
The fall is not as steep and pronounced as for e.g. 
IGR~J17511--3057 \citep{Ibragimov11,Falanga11}, IGR~J00291+5934 \citep{Falanga05b}, or  SAX J1808.4--3658 \citep{Cui98,Ibragimov09}, 
but rather similar to the behavior observed in XTE J1751--305 \citep{Gierlinski05} and XTE J1807--294 \citep{Chou08}. 
The time lags here, however, are rather small  and seem to saturate near 10 keV at $\sim -60\mu$s.

We note that the saturation occurs very close to the energy where thermal emission ceases, thus supporting the interpretation that the time lags are caused by a combination of the Doppler effect and 
different emission patterns of the thermal emission and the Comptonization tail \citep{Gierlinski02,PG03}.

\subsection{Spectrum of the pulsed emission and pulsed fraction}
\label{sec:pulsefrac}

The pulsed spectrum and pulsed fraction (defined as pulsed flux/total flux) as a function of energy provide important diagnostic parameters for constraining the parameter space in theoretical modeling \citep[see e.g.][]{Viironen2004}. 
We derived the pulsed fluxes ($\sim$3--150 keV) from observations with the (imaging) ISGRI instrument and 
the non-imaging \rxte/PCA and HEXTE instruments overlapping in time with the \Integ\ observations.
The total fluxes ($\sim$0.7--300 keV) were determined from instruments with imaging capabilities, namely \swift\ XRT, and JEM-X and ISGRI aboard \Integ.

Firstly, we derived the pulsed excess counts (=counts above DC level) in a given energy band by fitting a truncated Fourier series, using only the fundamental and two harmonics, to the measured pulse phase distribution.
In the case of the PCA data, only data in the $\sim$3--32 keV band were used because data below 3 keV and above 32 keV were excluded owing to calibration uncertainties. Furthermore, the pulsed excess counts were converted to photon flux values (ph cm$^{-2}$ s$^{-1}$ keV$^{-1}$)
in a forward spectral-folding procedure assuming an underlying power-law model and taking into account the different exposure times (see Sect. \ref{sec:time_data}) and energy responses (all three PCU layers were included) of the five active PCUs.
We kept the absorbing hydrogen column-density fixed to $0.96\times 10^{22}$ cm$^{-2}$, a value obtained from a broad-band (0.7--300 keV) fit of the total emission spectrum adopting the thermal Comptonization model {\sc compps} using data from \swift/XRT, \rxte/PCA, and
\Integ/JEM-X and ISGRI in the overlapping (contemporaneous) data period. This value differs  from the value $2.87\times 10^{22}$ cm$^{-2}$ reported by \citet{Torres11ATel3638} using {\em Chandra} HETGS MEG and HEG data over the 1--7 keV energy interval adopting a power-law model. The difference can easily be explained because of the usage of different energy intervals and different (assumed) underlying photon spectral models.

\begin{figure}
\begin{center}
\includegraphics[width= 0.85\columnwidth]{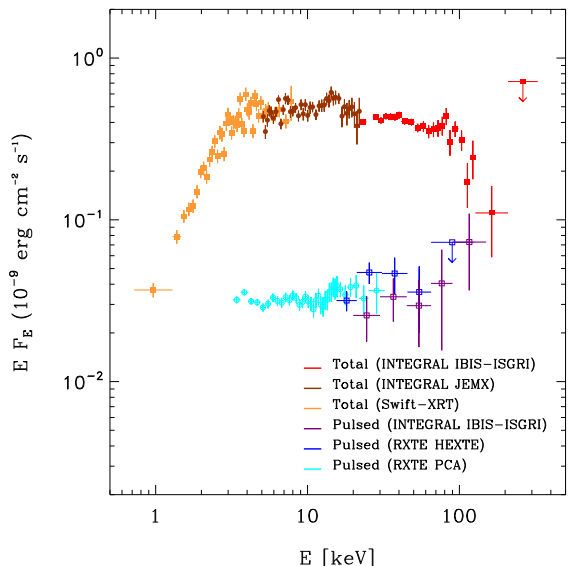}
{\caption{Unabsorbed total ($\sim$0.7--300 keV) and pulsed
($\sim$3--150 keV) unfolded spectra of \igr\ combining measurements derived from the \rxte\ PCA, HEXTE, and 
\Integ\ ISGRI data for the pulsed part and \swift\ XRT, \Integ\ JEM-X, and ISGRI data for the total part.
\label{fig:hea_spc}} }
\end{center} 
\end{figure}

We obtained an unabsorbed 2--10 keV pulsed flux of $5.12(7)\times 10^{-11}$ erg cm$^{-2}$ s$^{-1}$ and a photon power-law index of $1.968\pm0.007$ (reduced $\chi^2$ of the fit over the 3.2--32.2 keV band was 26.84/(36$-$2))  using PCA data collected in the \Integ\ overlapping data period. The power-law index we derived for the pulsed emission is compatible, using the same energy band and observation period, with the total emission spectrum, $\Gamma = 1.93\pm0.01$, which is also within the range of 1.8--2.0 reported by \citet{Papitto11}.
The best-fit \compps\ model to the pulsed emission and the total emission spectra were also found to agree, the same finding also being clearly visible in Fig. \ref{fig:hea_spc}, suggesting that there is an energy-independent pulsed fraction over the PCA band-pass.

For HEXTE data ($\sim$15--120 keV), we employed an equivalent method, where the number of pulsed excess counts in a certain energy band (for \rxte\ observations within the \Integ\ overlapping data period) were divided by its effective sensitive area assuming a power-law model of index 2, taking into account the energy response and dead-time corrected exposure time (see Sect. \ref{sec:time_data}) of detector cluster-0.
Finally, the pulsed ISGRI excess counts ($\sim$20--150 keV) of \igr\/ were converted to flux values by adopting the method outlined in Section 3.4 of \citet{Kuiper2006}.

\begin{figure}[t] 
\begin{center}
\includegraphics[width= 0.85\columnwidth]{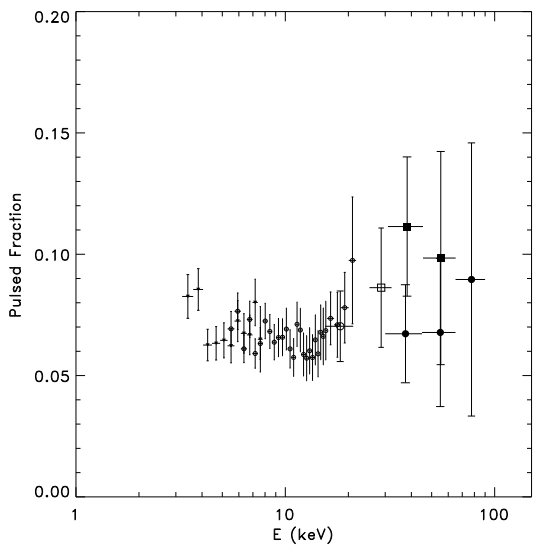}
 {\caption{The pulsed fraction (=pulsed flux/total flux) of \igr\ based on the pulsed/total flux measurements from \rxte/PCA/HEXTE and \Integ/ISGRI (pulsed) and \swift/XRT and \Integ/JEM-X/ISGRI (total). 
 The use of different symbols indicate that the ratios are based on different instrument combinations in overlapping energy bands.
 \label{fig:pulsedfraction}} }
 \end{center} 
\end{figure}

The (unabsorbed) PCA, HEXTE, and ISGRI pulsed flux measurements are shown in Fig.~\ref{fig:hea_spc} ($E F_E$ spectral representation), along with the (unabsorbed) total flux measurements derived for the same data period. From these pulsed and total flux measurements, we can easily derive the pulsed fraction as a function of energy. This is shown in Fig.~\ref{fig:pulsedfraction}. The pulsed fraction is consistent with being constant i.e. energy independent and has a typical value of 6--7\%. This behavior is in glaring contrast to that shown by e.g. IGR~J00291+5934 and IGR~ J17511--3057 \citep[see][respectively]{Falanga05b,Falanga11}, where clear energy-dependent variations can be seen. 
From the physical point of view this means that the emission pattern from the hotspot does not  change with energy. 
Because high-energy photons have undergone more scatterings in the hot electron slab than the low-energy photons, 
this means that for all observed energies the number of scatterings is rather large and therefore 
the emission pattern is similar \citep{Viironen2004}.

\section{X-ray bursts} 
\label{sec:burst} 

 \subsection{Data} 
\label{sec:burst_data} 

The imaging capabilities of \Integ\ allow us to separate the X-ray bursts of \igr\ from the bursts of other nearby sources,
which is impossible with  the \rxte/PCA camera.
Our analysis shows that among the several bursts detected by the PCA  only two are from the \igr\ source (see Table~\ref{tab:burst}). \citet{Chakraborty2012} reported 12 bursts identified in the \rxte\ datasets. 
They classified these bursts as bright, medium, and small type-I X-ray bursts. 
Using the imaging \Integ/JEM-X capability during the simultaneous \Integ/JEM-X and \rxte/PCA data, 
we were able to  associate the medium bursts with 1A1742--294 and the small bursts with SLX 1744--300/299. 
The only two bright type-I X-ray bursts were confirmed to have originated from \igr, a result that we consider in greater detal in this work.

For the \rxte\ detected bursts, where not explicitly stated, the data analysis procedures are as in \citet{Galloway2008}. 
We re-fit the burst spectra over the energy range 2.5--20 keV using the revised PCA response matrices, v11.7\footnote{http://www.universe.nasa.gov/xrays/programs/rxte/pca/doc/rmf/pcarmf-11.7 } and adopted the recommended systematic error of 0.5\%. The fitting was undertaken using {\sc xspec} version 12.7.0 \citep{Arnaud96}. To accommodate spectral bins with low count rates, we adopted Churazov weighting. The \rxte/PCUs are subject to a short ($\approx 10\ \mu$s) interval of inactivity following the detection of each X-ray photon. This ``dead-time'' reduces the detected count rate below the rate of photons incident on the detector (by an estimated 3\% for an incident rate of 400~count~s$^{-1}$~PCU$^{-1}$). We corrected each spectrum for dead-time by calculating an effective exposure, depending upon the measured count rate, which takes into account the dead-time fraction. The peak observed count-rate for burst  number 2 was 2920~count~s$^{-1}$~PCU$^{-1}$ (with data from PCU2 only), giving a dead-time correction of 7.7\%\footnote{ http://heasarc.nasa.gov/docs/xte/recipes/pca\_deadtime.html}. 
For bursts for which suitable data modes were available, time-resolved spectra in the range 2--60~keV covering the burst duration were extracted on intervals as short as 0.25 s during the burst rise and peak. The bin size was gradually increased into the burst tail to maintain roughly the same signal-to-noise level. A spectrum taken from a 16-s interval prior to the burst was adopted as the background. 

For the \Integ\ detected bursts, we fit the spectra collected within 4 s time bin intervals. 
As background, we used the whole-pointing burst-subtracted spectrum. 
None of the \Integ/JEM-X detected bursts from \igr\ had been observed with the ISGRI instrument at higher energy (18--40 keV).

\begin{table*} 
\caption{Burst characteristics observed with \rxte/PCA, \swift/XRT, and \Integ/JEM-X}
\centering   
\begin{tabular}{lccccccc} 
\hline 
 \#  & Observatory& Burst $T_{\rm start}$ & $F_{\rm pers, bol}$\tablefootmark{a}& 
 $F_{\rm peak}$\tablefootmark{b} & $f_{\rm b}$\tablefootmark{c} & $\tau_{\rm b}$\tablefootmark{d} & 
 $kT_{\rm bb,peak}$\tablefootmark{e}\\ 
 & &(UT) & $10^{-9}$ erg cm$^{-2}$ s$^{-1}$ & 
 $10^{-9}$erg cm$^{-2}$ s$^{-1}$ & $10^{-7}$erg cm$^{-2}$& s & keV\\
\hline 
\noalign{\smallskip}  
1 & {\em INTEGRAL}&2011-08-14 03:21:26 & $1.95\pm0.15$ & $56.0\pm2.2$ & $6.0\pm0.8$ & $10.7\pm1.5$& $2.0\pm0.5$\\
2 & {\em RXTE}&2011-08-16 15:21:45 & $2.05\pm0.70$ & $48.8\pm1.5$ & $4.3\pm0.4$ & $8.9\pm0.8$ & $2.69\pm0.07$\\
3\tablefootmark{f} &{\em Swift} & 2011-08-18 11:23:52 & $2.02\pm0.80$ & $49.0\pm1.9$ & $4.0\pm0.6$ & $8.2\pm1.3$& $1.9\pm0.3$\\
4 &{\em INTEGRAL} & 2011-08-19 03:24:26 & $1.95\pm0.11$ & $49.0\pm2.2$ & $4.5\pm0.8$ & $9.2\pm1.7$& $2.0\pm0.5$ \\
5 & {\em INTEGRAL} & 2011-08-19 19:59:09 & $1.89\pm0.11$ & $67.8\pm2.2$ & $4.7\pm0.8$ & $6.9\pm1.2$& $2.3\pm0.5$ \\
6\tablefootmark{g} & {\em RXTE} & 2011-08-20 14:10:08 & $1.85\pm0.20$ & $27.0\pm0.2$ & $5.56\pm0.04$ & $20.6\pm0.2$ & $2.13\pm0.01$ \\
7 & {\em INTEGRAL} & 2011-08-27 00:35:40 & $1.42\pm0.80$ & $61.0\pm2.2$ & $5.1\pm0.8$ & $8.4\pm1.3$& $2.2\pm0.5$ \\
8 &{\em Swift} & 2011-08-28 10:13:22 & $1.25\pm0.10$ & $42.0\pm1.9$ & $3.9\pm0.7$ & $9.3\pm1.7$& $2.3\pm0.3$ \\
\noalign{\smallskip}  
\hline  
\noalign{\smallskip}  
\label{tab:burst} 
 \end{tabular} 
\tablefoot{ 
\tablefoottext{a}{Pre-burst unabsorbed flux in the 0.1--300 keV energy range.}
\tablefoottext{b}{Burst peak flux in the 0.1--40 keV energy band.}
\tablefoottext{c}{Burst fluence in the 0.1--40 keV energy band.}
\tablefoottext{d}{Effective duration $\tau_{\rm b}=f_{\rm b}/F_{\rm peak}$.}
\tablefoottext{e}{Burst peak temperature.}
\tablefoottext{f}{Burst also observed in the \Integ/JEM-X data.}
 \tablefoottext{g}{Burst analysis using {\em RXTE}/PCA standard two mode data with16 sec time resolution, therefore, $F_{\rm peak}$ and $kT_{\rm bb,peak}$ are more accurately interpreted as limits, owing to the low time resolution of the source data.}
}
\end{table*}
   
 \subsection{Burst properties} 
\label{sec:burst_prop} 
   
In Table~\ref{tab:burst}, we report the key measurable parameters for the
bursts observed from \igr.
Thermonuclear (type-I) X-ray bursts are produced by unstable burning of accreted
matter on the NS surface. The spectrum from
a few keV to higher energies can usually be well-described 
as a blackbody with temperature $kT_{\rm bb}\approx$1--3 keV. 
The energy-dependent decay time of these bursts is
attributed to the cooling of the NS photosphere and results in a gradual
softening of the burst spectrum \citep[see][for a review]{Lewin1993,Strohmayer2006}. 

We defined the burst start time as the time at which
the X-ray intensity of the source first exceeded 25\% of the burst peak
flux (above the persistent intensity level). The time-resolved spectral
analysis of the eight bursts (two \rxte, five \Integ, and two \swift)\footnote{Burst  number 3 reported in Table~\ref{tab:burst} was observed simultaneously 
with \Integ/JEM-X, as well as with \swift/XRT.} was carried out using the 
\rxte/PCA and \Integ/JEM-X data in the 2.5--20~keV and
3--20~keV bands, respectively. From these analyses, we determined the
bursts' peak fluxes, temperatures, and radii (see Table
\ref{tab:burst}).  We fitted each burst spectrum by a simple
photoelectrically absorbed blackbody model,
\bb. The neutral absorption column density \nh\ was fixed at the value of
 $1.2\times10^{22}$ cm$^{-2}$ in all fits. However, we
checked that leaving \nh\ free 
did not significantly affect the results. We extrapolated  the
unabsorbed fluxes to the 0.1--40~keV band by
generating dummy responses ({\sc xspec} version 12.7.0). This is
justifiable for the data because the blackbody temperature is well
inside the spectral bandpass. 
The inferred \bb\ peak temperature, $kT_{\rm bb, peak}$,
and unabsorbed bolometric peak flux are also reported in Table \ref{tab:burst}.

\begin{figure}[t] 
\centering 
\includegraphics[width= 0.90\columnwidth]{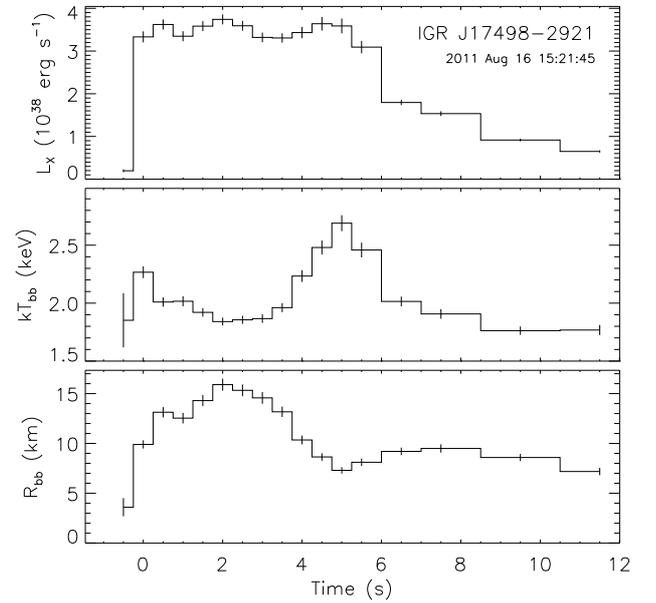}
{\caption[]{Representative time-resolved spectroscopic results from the brightest burst (PRE burst number 2 from Table \ref{tab:burst}) 
seen by {\it RXTE}\/ from \igr. We show the inferred bolometric luminosity inferred by integrating the best-fit blackbody model ({\it top panel}), 
the blackbody temperature ({\it middle panel}),  and the effective blackbody radius ({\it bottom panel}), assuming a distance of 8~kpc. 
\label{fig:specburst}}}
\end{figure} 

When a burst undergoes a photospheric-radius expansion (PRE), the
source distance can be determined based on the assumption that the
bolometric peak luminosity is saturated at the Eddington limit,
$L_{\rm Edd}$ \citep[e.g.,][]{Lewin1993}. 
The only burst in our sample that exhibited PRE was burst number 2 
(see Fig.~\ref{fig:specburst} for the time-resolved spectral parameters). 
Assuming a bolometric peak luminosity equal to the Eddington value for a He X-ray burst 
\citep[$L_{\rm Edd}\approx\,3.8\times 10^{38}$ ergs$^{-1}$, as empirically derived
by][]{Kuulkers2003}, we obtained the source distance of $d\approx8.1$ kpc. 
For comparison, the theoretical value of the distance \citep[e.g.,][]{Lewin1993} found 
by assuming a He atmosphere and canonical NS parameters (1.4 solar mass and radius of 10 km) is $\sim6.6$ kpc. 
Using the brightest burst (number 5) peak flux reported in Table~\ref{tab:burst}, the empirical distance is $\sim6.8$ kpc. 
However, because the JEM-X statistic are too poor to perform a time-resolved spectral analysis to detect a PRE, 
we considered $d\approx$8~kpc to be a fiducial distance derived from burst number 2.

This burst had a temperature of about 2.7 keV at the touchdown, 
where the blackbody radius was 7.3 km.  
After the touchdown at the cooling track, the radius slightly increased and then decreased. 
This evolution of the apparent radius was probably caused by variations of the color-correction factor, 
$R_{\rm bb}\propto f_{\rm c}^{-2}$. 
The amplitude of the variations is consistent with the theoretical models \citep{Suleimanov11}, which predict 
about 20\% changes in $f_{\rm c}$, when the flux drops by a factor of three from the peak (Eddington) value. 

We calculated the burst fluence, $f_{\rm b}$, by integrating the flux over the burst. We estimated the additional contribution to the fluence after the last time bin for which the flux could be measured, by integrating an exponential curve fitted to the last few flux points to infinity. This additional contribution was adopted as the error in the fluence for the cases where it exceeded the statistical error in the fluence calculated from the measured fluxes (e.g. burst number 2). The fluences and peak fluxes for the bursts were roughly constant with time, with $f_{\rm b} \approx 5\times 10^{-7}\, {\rm erg\,cm^{-2}}$, and $F_{\rm peak}\approx 5 \times 10^{-8}\, {\rm erg\,cm^{-2}\,s^{-1}}$ (excluding burst number 6, for which high-time resolution data were not available, thus the peak flux was underestimated). We also calculated the effective burst duration $\tau_{\rm b}=f_{\rm b}/F_{\rm peak}$, which was  roughly constant throughout the outburst, at $\approx$6--10~s (except burst number 6). 
The burst rise time was generally shorter than the spectral bin size, which at its shortest was 0.25~s (for burst number 2).

Analysis of a larger sample of type-I bursts observed by \rxte\ indicate that bursts with short rise times and $\tau_{\rm b}<10$ s are generally consistent with hydrogen-deficient fuel \citep[][Fig. 21]{Galloway2008}. Thus, on the basis of the burst lightcurves, we expect that this was also the case for the bursts from \igr. 

 \subsection{Physical implications} 
\label{sec:burst_physics}

We now attempt to place more precise constraints on the burst fuel and ignition conditions. At the inferred distance of 8~kpc, the bursts occurred at persistent luminosities between  $(0.96-1.6)\times10^{37}$ erg s$^{-1}$ (see Table~\ref{tab:burst}), 
or $\approx(2.5-4.1)\%$ of $L_{\rm Edd}$ (using $L_{\rm Edd}\approx\,3.8\times 10^{38}$ erg s$^{-1}$).  
The local accretion rate per unit area for the pre-burst emission, $L_{\rm pers}$, is then
given by $ \dot m = L_{\rm pers} (1+z) (4\pi R^2(GM/R))^{-1}$,
i.e. $\dot m \sim (5.3-8.8)\times10^3 $g cm$^{-2}$ s$^{-1}$.  We used here the gravitational redshift $1+ z = 1.31$ for the canonical NS mass, $M=1.4 M_{\odot}$, and radius, $R=10$ km. 

Bursts numbers 3--6 occurred during uninterrupted observations with {\it INTEGRAL}, so there is little uncertainty about the burst recurrence times for these bursts. The burst number 6 was detected only with \rxte, and occurred during a \rxte/PCA observation that started shortly before the end of our INTEGRAL target of opportunity observation. We calculated the usual ratio of the integrated persistent flux to the burst fluence, $\alpha=\tau_{\rm rec}F_{\rm pers}/f_b$, where $F_{\rm pers}$ is the bolometric persistent flux between each burst and $\tau_{\rm rec}$ is the recurrence time. 
The  $\alpha$-values for bursts  4--6 were consistent with a constant value of $\approx250$. This implies a nuclear energy release  
$Q_{\rm nuc}=m_{\rm p} c^2z/\alpha\approx1.2$~MeV~nucleon$^{-1}$. The usual approximation for $Q_{\rm nuc} \approx 1.6+4\left<X\right>$~MeV~nucleon$^{-1}$ (including losses owing to neutrino emission following \citealt{Fujimoto1987}, 
where $\left<X\right>$ is the mean H-fraction of the fuel layer, see e.g. \citealt{Galloway2004} and references therein), 
implies that the measured $Q_{\rm nuc}$ is below that of even pure He fuel (i.e. $\left<X\right>=0$). 

This otherwise unphysical situation may be explained by the anisotropy of the burst and persistent emission. Implicit in the calculation of $\alpha$ is a term $\xi_{\rm p}/\xi_{\rm b}$, which describes the relative beaming of the persistent and burst emission \citep[e.g.,][]{Fujimoto1988,Lapidus1985}. The different geometries of the persistent and burst emission imply that the degree of anisotropy of each, parametrized as $\xi_{\rm b,p}$ (where $\xi<1$ implies that the emission is beamed preferentially towards the observer) may be different. The true value of $Q_{\rm nuc}$ may be larger than that inferred from the $\alpha$ measurement, if $\alpha$ is in turn exaggerated owing to the persistent emission being beamed towards the observer to a greater extent than the burst (i.e. $\xi_{\rm p}<\xi_{\rm b}$). \citet{Fujimoto1988} find a minimum value of the ratio $\xi_{\rm p}/\xi_{\rm b}=0.75$ for a system inclination of zero. This is sufficient to bring the inferred value of $Q_{\rm nuc}$ close to 1.6, the expected value for pure He-fuel. We thus infer that the bursts from \igr\ are powered by H-poor material, but also that the system is observed at low inclination.
The fuel composition at ignition may be modified by steady H-burning prior to the burst. The H-burning proceeds via the $\beta$-limited hot CNO cycle, and exhausts the available hydrogen fuel in a time of $t_{\rm burn} = 11 (Z/0.02)^{-1} (X_0/0.7)$~hr \citep[e.g.,][]{Galloway2004}, where $Z$ is the CNO mass fraction, and $X_0$ the H-fraction of the accreted fuel. Thus, the measured recurrence times are sufficiently long  to ensure that even for a solar hydrogen mass fraction in the accreted fuel, steady burning can exhaust the hydrogen at the base, producing the requisite short, He-like burst profiles and high $\alpha$-values.

The observed energies of the bursts allow us to estimate the ignition
depths. The measured fluences of the bursts are
$f_{\rm b}=(3.9-6.0)\times 10^{-7}\ {\rm erg\ cm^{-2}}$, corresponding to a
net burst energy release $E_{\rm burst}=4\pi
d^2f_{\rm b}=(3.0-4.6)\times 10^{39}\ (d/8\ {\rm kpc})^2\ {\rm erg}$. The
ignition depth is given by $ y_{\rm ign} = E_{\rm burst} (1+z)(4\pi
R^2Q_{\rm nuc}/m_{\rm p})^{-1}$. For the inferred pure-helium composition at ignition (i.e. $\langle
X\rangle=0$) the column depth varies little from burst to burst, in the
range $y_{\rm ign}\approx(1.9-3.0)\times 10^{8}\ {\rm g\ cm^{-2}}$.

The theoretical recurrence times for all the bursts, calculated with the equation $\tau_{\rm rec} = (y_{\rm ign}/\dot m)(1+z)$ and assuming the values for $y_{\rm ign}$ and $\dot m$ reported above, are in the range of 8--20 hr. In particular, for the bursts numbers 3--6 we obtained from this equation recurrence times of Ê$9.5^{+4.5}_{-3}$, $10.3^{+4.5}_{-3}$, and Ê$12.5^{+2.5}_{-2}$ hr, respectively. These are significantly shorter than those measured from the observations of 16.01 hr, 16.58 hr, and 18.18 hr (we note that bursts numbers 3--6 occurred during uninterrupted observations). A similar result was obtained when the rotational evolution of a weakly magnetized neutron-star model \citep[][solar CNO metallicity and a hydrogen mass fraction of 0.7 are assumed]{Cumming2000} was used instead of the simpler equation above. According to this model, the recurrence time corresponding to a bolometric flux of $\sim 2\times10^{-9}$ erg cm$^{-2}$ s$^{-1}$ (a mean value between those of the three bursts ) would be $\sim$7~hr. In both cases, we thus obtained recurrence times substantially shorter than observed. The inferred low inclination from the burst energetics implies that the isotropic luminosity is overestimated by a factor of $\sim2$, which would reduce the inferred accretion rate by the same factor, and give a recurrence time of just over 12~hr. Although this remains somewhat shorter than predicted, slight decreases in either the CNO or hydrogen mass fraction in the accreted fuel could bring the predicted rates back into agreement with those observed. We defer a more detailed treatment of the burst energetics to a future paper.
 
\section{Summary} 
\label{sec:summary} 

We have analyzed simultaneous \Integ, \rxte, and \swift\ observations to study the
broad-band spectrum and timing behavior of \igr. Using all these data, we also studied the outburst profile. The  broad-band average spectrum is well-described by thermal Comptonization with an electron temperature of $\sim$50 keV, seed photon temperature of $\sim$1 keV, and Thomson optical depth $\taut\sim1$ in a slab geometry. In addition this object shows remarkable spectral stability during the outburst, as marked by  constant plasma and seed photon temperatures at a constant scattering optical depth.  

We have shown that the coherent pulsation can be tracked with the HEXTE and ISGRI instruments 
up to $\sim$65 keV.  The pulsed fraction was found to be constant, i.e., energy-independent, and has a typical value of 6--7\%. The nearly sinusoidal pulses show soft lags beyond $\sim 4$ keV (hard X-ray photons arrive earlier than the soft ones) that seems to saturate near 10 keV at 
$\sim-60\mu$s. The fall is not as steep and pronounced as for example found for either IGR~J17511--3057 or IGR~J00291+5934. 

Using all observations by \Integ, \rxte, and \swift, we have collected the most comprehensive set of X-ray bursts observed from \igr, which has allowed us to determine the recurrence time as a function of the accretion rate and the ignition depth. The short burst profiles indicate that there is hydrogen-poor material in the process of ignition, which suggests either that the accreted material is hydrogen-deficient or that the CNO metallicity is up to a factor of two times solar. However, the variation in the burst recurrence time as a function of the accretion rate  (inferred from the X-ray flux) is much smaller than predicted by helium-ignition models.
   
\begin{acknowledgements} 
The authors thank Chris Winkler and the \Integ\ team for the rapid scheduling of the observations of \igr\ shortly after the onset of its outburst. 
JP acknowledges financial support from the Academy of Finland grant 127512. 
MF, JP, and DG thank the International Space Science Institute (ISSI) in Bern for hosting an International Team on type-I X-ray bursts. 
\end{acknowledgements} 
 

\end{document}